\documentclass[aps,prb,twocolumn,superscriptaddress,showpacs]{revtex4-1}

\usepackage{graphicx,psfrag}
\usepackage[latin1]{inputenc}
\usepackage{natbib}
\usepackage{hyperref}
\usepackage{amsmath, amsthm, amssymb}
\usepackage{color}
\usepackage{soul}
\usepackage{comment}
\usepackage{hyperref}
\usepackage{amsfonts}
\usepackage{fancyhdr}

\newcommand{\notes}[1]{}

\newcommand{\beq}{\begin{equation}}
\newcommand{\eeq}{\end{equation}}
\newcommand{\beqnn}{\begin{equation*}}
\newcommand{\eeqnn}{\end{equation*}}
\newcommand{\beqas}{\begin{eqnarray*}}
	\newcommand{\eeqas}{\end{eqnarray*}}
\newcommand{\beqa}{\begin{eqnarray}}
\newcommand{\eeqa}{\end{eqnarray}}

\begin{document}
	
\title{Magnetic Droplet Solitons}
	
\author{Ferran Maci\`a}
\affiliation{Dept.\ of Condensed Matter Physics, University of Barcelona, 08028 Barcelona, Spain}
\affiliation{Institute of Nanoscience and Nanotechnology (IN2UB),
University of Barcelona, 08028 Barcelona, Spain}
\author{Andrew D. Kent}
\affiliation{Center for Quantum Phenomena, Department of Physics, New York University, New York, New York 10003 USA}
\date{\today}

\begin{abstract}
Magnetic droplet solitons are dynamical magnetic textures that form due to an attractive interaction between spin waves in thin films with perpendicular magnetic anisotropy. Spin currents and the spin torques associated with these currents enable their formation as they provide a means to excite non-equilibrium spin wave populations and compensate their decay. Recent years have seen rapid advances in experiments that realize and study magnetic droplets. Important advances include the first direct x-ray images of droplets, determination of their threshold and sustaining currents, measurement of their generation and annihilation time and evidence for drift instabilities, which can limit their lifetime in spin-transfer nanocontacts. This article reviews these studies and contrasts these solitons to other types of spin-current excitations such as spin-wave bullets, and static magnetic textures, including magnetic vortices and skyrmions. Magnetic droplet solitons can also serve as current controlled microwave frequency oscillators with potential applications in neuromorphic chips as nonlinear oscillators with memory.
\end{abstract}
\maketitle

\section{Introduction}
Magnetic droplet solitons are a localized region of highly excited spin waves that form in thin films with perpendicular magnetic anisotropy. Until recently, these objects while intriguing---and potentially useful in information processing---were not possible to realize and study experimentally. This is because they were predicted to occur in materials without magnetic damping and to require a non-equilibrium spin wave (or magnon) population~\cite{Ivanov1977,Kosevich1990}. All magnetic materials have damping so that spin waves decay toward an equilibrium population set by the temperature. 

The discovery of spin torques changed this situation~\cite{Slonczewski1989,slonczewski1996current,berger1996emission}.
A flow of spin angular momentum can compensate damping in a magnetic material and also create spin waves. 
Spin currents can thus create the conditions for the formation of a ``dissipative droplet solitons,'' \cite{Rippard2010,mohseni2011high,Mohseni2013} which were predicted to occur in a nanocontact to a ferromagnetic layer~\cite{Hoefer2010}. Spin polarized current flow through the nanocontact leads to spin-wave excitations that localize in the nanocontact region forming a droplet soliton or droplet, for short.

There are a number of important fundamental characteristics of such droplets. First, they are generated at a threshold current that is determined by the magnetic damping and magnetic field. Second, their spin-precession frequency is less than the lowest propagating spin-wave modes in the film, the ferromagnetic resonance frequency. Third, their sustaining current can be lower than their generating current, that is, once generated they can be sustained at a lower current. Further, they have unique collective dynamics, including low frequency motion, drift instabilities and multiple modes.

Interest in applications is related to their dynamical characteristics. Spin precession in a nanocontact can lead to resistance oscillations associated with magnetoresistance of the contact, denoted a spin transfer nanocontact oscillator (STNO). An STNO containing a droplet is thus a current controlled oscillator with oscillation frequencies in the GHz range. These are nonlinear oscillators in that droplet characteristics are a nonlinear function of parameters, such as the current and magnetic field. Further, their particle-like nature, which includes the possibility of droplet interaction and self modulation, together with hysteresis in their response, make these oscillators of interest in neuromorphic computing~\cite{Hoppensteadt1999,Macia2011,Locatelli2014,Romera2018}, such as for implementing reservoir computing~\cite{Torrejon2017}.

\section{Droplet Solitons}
\begin{figure}
	\begin{center}
		\includegraphics[width=\columnwidth]{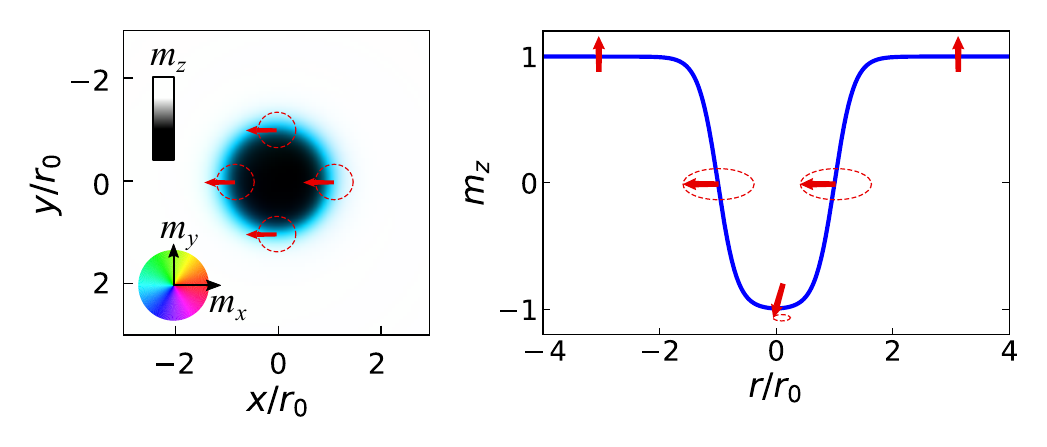}
		\caption{Schematic of a droplet soliton. Left: droplet in a nanocontact with radius $r_0$. The color indicates the direction of magnetization. Right: Profile through the droplet core. Spins precess about the anisotropy field, i.e. normal to the film plane. At the droplet boundary the precession amplitude is maximum, as indicated by the red arrows.}
		\label{Fig:DS}	
	\end{center}
\end{figure}

A droplet is a nearly circular region of suppressed z-component of magnetization in which the spins precess about the z-axis, in the x-y plane (Fig.~\ref{Fig:DS}). At its center the spin-precession amplitude is minimal and the magnetization is nearly reversed (i.e. $m_z\simeq-1$), while at the droplet boundary ($m_z\simeq 0$) the spin-precession amplitude is maximal.  
A profile indicating the spin precession is shown in Fig.~\ref{Fig:DS}(b).

\begin{figure}[b]
	\begin{center}
		\includegraphics[width=\columnwidth]{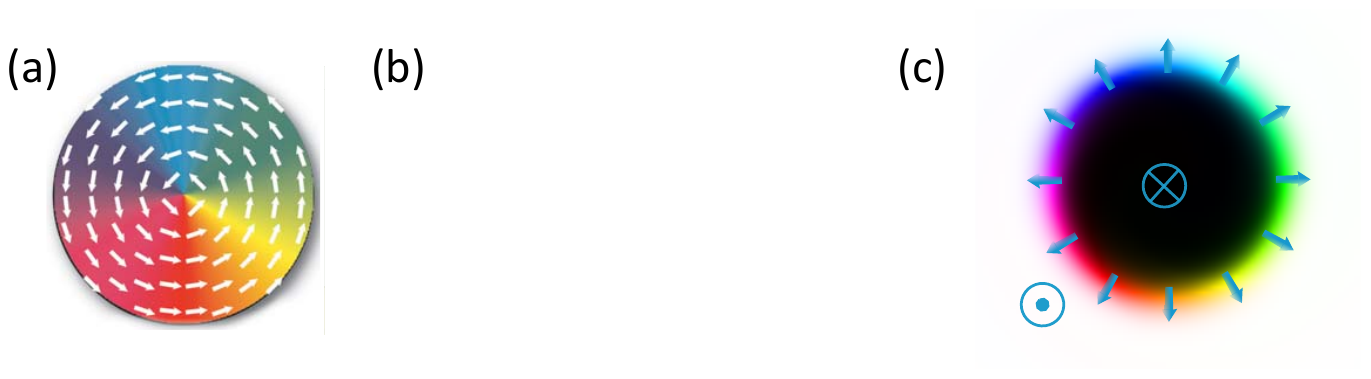}
		\caption{(a) A magnetic vortex. (b) A magnetic skyrmion. In contrast to a droplet soliton these are stable or metastable objects that can have static spin configurations. (c) A topological magnetic droplet. The spins at the droplet boundary oscillate from a Bloch configuration (radial direction, shown) to a N\'eel configuration (spins in the azimuthal direction).}
		\label{Fig:Vortex}	
	\end{center}
\end{figure}

A droplet is thus a dynamical object in contrast to static magnetic solitons, such as magnetic vortices and skyrmions shown in Figs.~\ref{Fig:Vortex}(a) and (b). The latter are generally stable or metastable static spin configurations in a magnetic material. Magnetic vortices can form in disks to minimize the dipolar magnetic interactions at the expense of exchange interactions~\cite{Shinjo2000}. They can also be formed in nanocontacts, as a current creates a circular `Oersted' magnetic field that can favor this configuration~\cite{Manfrini2013}. Skyrmions are favored in the presence of anisotropic exchange interactions known as Dzyaloshinskii-Moriya interactions~\cite{Fert2017}, although they can also be stabilized by dipolar interactions~\cite{Jiang2015}. In the latter case, they tend to be very large objects ($\sim \mu$m). Both vortices and skyrmions are topological objects, meaning that they cannot be removed by a continuous deformation of the magnetization. Droplets, in their most basic form, are non-topological. However, topological droplets are also possible, as illustrated in Fig.~\ref{Fig:Vortex}(c)~\cite{Kosevich1990,Zhou_natComm_2015,Liu_prl_2015,Carpentieri2015,Statuto2018}. 
Micromagnetic modeling has shown that the formation of topological droplets can be favored for certain initial magnetization states and current pulse rise times~\cite{Statuto2018}. In these types of solitons the boundary spins oscillating between Bloch and N\'eel configurations while maintaining a fixed topological charge. In common with static solitons, droplets can be metastable and have an energy both to formation and annihilation, as analyzed recently~\cite{Chaves2020} and seen in the experiment, which we discuss further below.

\begin{figure}[t]
	\begin{center}
		\includegraphics[width=85mm]{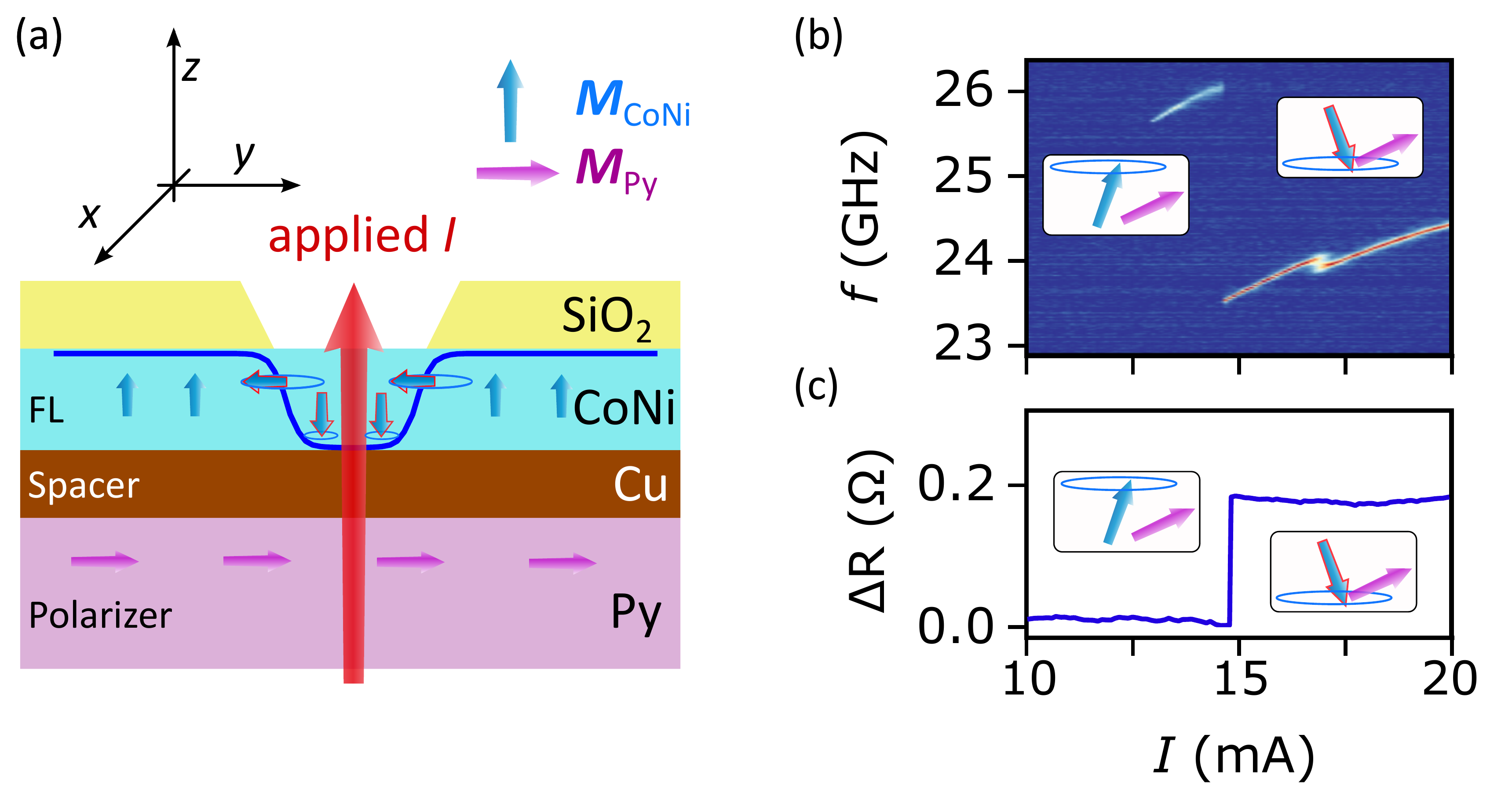}
		\caption{Nanocontact and its electrical characteristics. (a) Schematic of a spin transfer nanocontact with a droplet. An electrical current flows through a nanocontact to a thin ferromagnetic layer (the free layer, FL) and spin-polarizing layer. The external magnetic field is applied perpendicular to the film plane (the z-direction). The droplet is the nearly reversed magnetization region with spins precessing in the x-y plane. Blue arrows correspond to the magnetization of the free layer (CoNi) and pink arrows to the magnetization of the polarizing layer (Py). (b) High-frequency spectra as a function of current of a 100-nm-diameter contact in an applied fields of $\mu_0H =$ 0.85 T. The peak in the spectral response at 26 GHz is close to the  FMR frequency. The abrupt decrease in peak frequency at 15 mA is consistent with droplet formation. 
		(c) Nanocontact resistance as a function of current. Droplet formation leads to a step increase in resistance.}
		\label{Fig:Nanocontact}	
	\end{center}
\end{figure}

As noted, the formation of a droplet requires a spin-transfer torque. A typical sample and experimental setup is shown schematically in Fig.~\ref{Fig:Nanocontact}(a). The sample consists of two ferromagnetic layers separated by a non-magnetic layer; current flow is vertical in this image, perpendicular to the plane of the layers. One of the ferromagnetic layers, the free layer, FL, (the magnetic layer that responds to the spin torques), has perpendicular magnetic anisotropy and the other ferromagnetic layer, the fixed or polarizer layer, PL, usually has in-plane magnetic anisotropy. This leads to an orthogonal alignment of the layer magnetizations in the absence of an applied field. The intervening non-magnetic layer is sufficiently thick to magnetically decouple the layers (i.e. $>3$ nm, set mainly by Ruderman-Kittel-Kasuya-Yosida interactions) yet thin enough that spin is conserved in electron flow between the magnetic layers (i.e., thinner than its spin-diffusion length $\simeq 100$ nm for Cu).

The contact is characterized electrically by measuring its resistance-current response and voltage noise spectra at fixed current and applied field. An example of the latter measurements are shown in Fig.~\ref{Fig:Nanocontact}(b). As the fixed layer can have an in-plane component of magnetization (even in relatively large perpendicular applied fields, $H<M_s$, $\mu_0 H \lesssim 1$ T, for a Ni$_{80}$Fe$_{20}$, Permalloy, Py, fixed layer), spins precessing in the droplet leads to resistance oscillations at the spin precession frequency due to the giant magnetoresistance effect. This frequency is less than the film's FMR frequency. There is thus a step-like decrease in the peak noise frequency  (Fig.~\ref{Fig:Nanocontact}(b), $I>15$ mA). A step increase in resistance occurs when the droplet forms (Fig.~\ref{Fig:Nanocontact}(c)), again associated with giant magnetoresistance. 

\section{Imaging Droplets}
X-ray imaging experiments have provided the first direct evidence for droplet formation in nanocontacts~\cite{Backes2015}. They have also allowed measurements of their spin profile~\cite{Backes2015,Chung2018}. These experiments use resonance x-ray magnetic circular dichroism to probe specific elements in the nanocontact with high spatial and even temporal resolution~\cite{Bonetti2015}. Experiments have used a scanning transmission x-ray microscope (STXM) at a synchrotron~\cite{Bonetti2015RSI} as well as x-ray holography to image droplets~\cite{Parra2018}.

\begin{figure}[t]
	\begin{center}
		\includegraphics[width=\columnwidth]{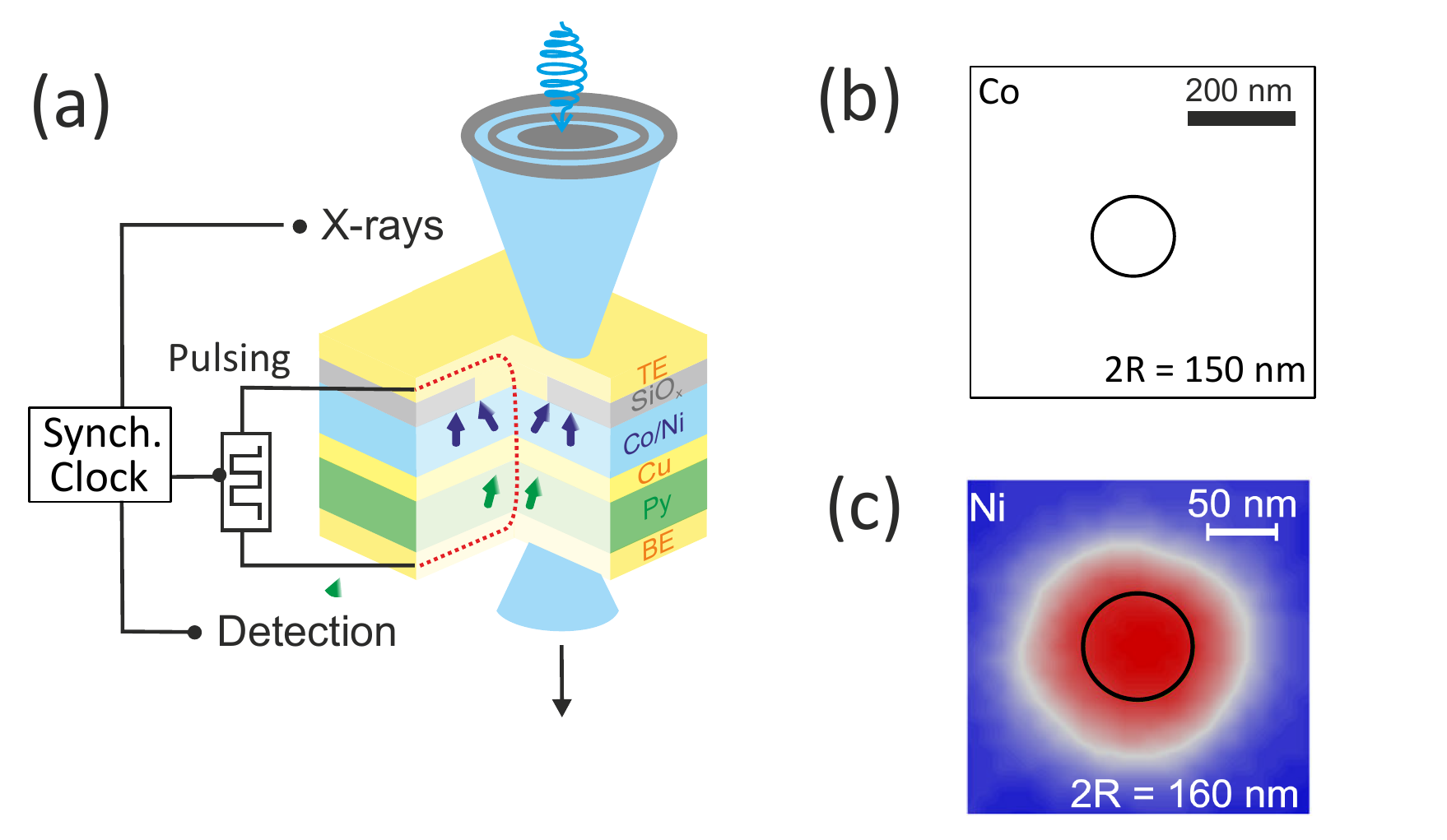}
		\caption{(a) Schematic of the STXM instrument and the 
			sample. The nanocontact is patterned on a SiN window to enable x-ray transmission. A Fresnel zone plate was used to focus the x-ray beam to
			a 35 nm spot, which was scanned across the area around the
			nanocontact, indicated as the yellow region contacting the Co/Ni
			layer through the SiO$_2$ dielectric, to acquire an image. 
			The x-ray detection was synchronized with the x-ray pulses from the
			synchrotron (rf clock) at 476.2 MHz. Spatial STXM images of the $M_z$ component of the Co  in (b) for an orthogonal sample Ref. \cite{Backes2015} and for Ni in (c) for an all perpendicular sample Ref.~\cite{Chung2018}}.
		\label{Fig:Images}	
	\end{center}
\end{figure}

Figure~\ref{Fig:Images}(a) shows an STXM experimental setup. The x-ray beam is perpendicular to the sample surface. Thus the absorption signal probes the perpendicular component of the magnetization ($m_z$), the envelope of the magnetic excitation. In Ref.~\cite{Backes2015} the x-ray energy was tuned to be resonant with the Co L$_3$ edge. As Co is only in the free layer (as the fixed layer is NiFe), the experiment is only sensitive to magnetization changes in the free layer. An image is acquired by scanning the sample in the x-ray spot, which is about 35 nm in diameter; one image can take about 20 minutes to acquire.

Figure~\ref{Fig:Images}(b) shows experimental results with a current above the threshold to nucleate a droplet. Red in the image corresponds to a reduced $m_z$. The reduced magnetization is consistent with droplet formation. A careful analysis of the absorption signal shows that it has a full width at half maximum of 175 nm, close to the nominal diameter of the nanocontact (150 nm). However, the reduction in magnetization at the center of the nanocontact corresponds to an angle of precession of 25 degrees, in contrast to the nearly complete magnetization reversal (180 degrees) expected in a nanocontact of this type~\footnote{Different nanocontact radii and and material parameters could lead to a droplet soliton profile 
where $m_z(0)$ is well above -1, i.e., where the magnetization in the nanocontact center need not be fully reversed}.

There are several possible reasons that a full reversal is not seen. First, as discussed in the sections below, the droplets are not stable in the nanocontact region and experience drift instabilities (see Sec.~\ref{Sec:Drift}), drifting out of the contact and then reforming. As the image averages over long time scales, times with and without a droplet will be averaged resulting on a lower average precession amplitude. Second, the nanocontact current is modulated on and off at a frequency of 1.28 MHz to increase the signal-to-noise ratio. The current pulse is thus on for 400 ns, which is comparable to the droplet nucleation time (see Sec.~\ref{Sec:timescales}, Fig.~\ref{Fig:Timescales}(a)). Thus a droplet may not fully nucleate in each current pulse cycle.

Attempts have been made to render droplets more stable to enable such static imaging experiments. One approach has been to use a spin polarizing layer with perpendicular magnetization~\cite{Chung2018}. The result of this study are shown in Fig.~\ref{Fig:GenAnn}(c). The droplet appeared to be about 3 times larger than the nanocontact size. This was ascribed to the Zhang-Li~\cite{ZhangLi2004} torque on the droplet boundary, which modifies not only the droplet effective size but also the threshold current~\cite{Albert2020}. However, it also could be associated with droplet dynamics, as the experiment also showed low frequency noise, as discussed further in Sec.~\ref{Sec:Drift}. The core of the droplet was found to be fully reversed in these experiments.

\section{Droplet generation and annihilation}
\label{Sec:GenAnn}
A critical value of the spin torque is required to generate a droplet, which translates into the existence of a critical or threshold current and field for droplet formation. However, experiments have shown the existence of current and field regions where droplets cannot be generated but can be sustained, indicating clearly that droplet states have hysteresis~\cite{Macia2014}.

In orthogonal magnetic anisotropy samples, at zero applied field, the FL and PL magnetizations are perpendicular, as shown in Fig.~\ref{Fig:Nanocontact}. In this case, current that is polarized in the magnetization direction of the PL creates a torque that averages to zero over one precessional cycle of the FL magnetization. As the magnetic field applied perpendicular to the layer plane increases, the PL magnetization tilts out of the film plane and this increases the current's spin polarization in the direction of the FL magnetization; eventually, the spin torque on the FL compensates the damping leading to droplet nucleation. Larger applied fields ultimately destabilize and annihilate droplets, favoring alignment of the two layers' magnetization in the direction of the applied field.

\begin{figure*}[t]
	\begin{center}
		\includegraphics[width=145mm]{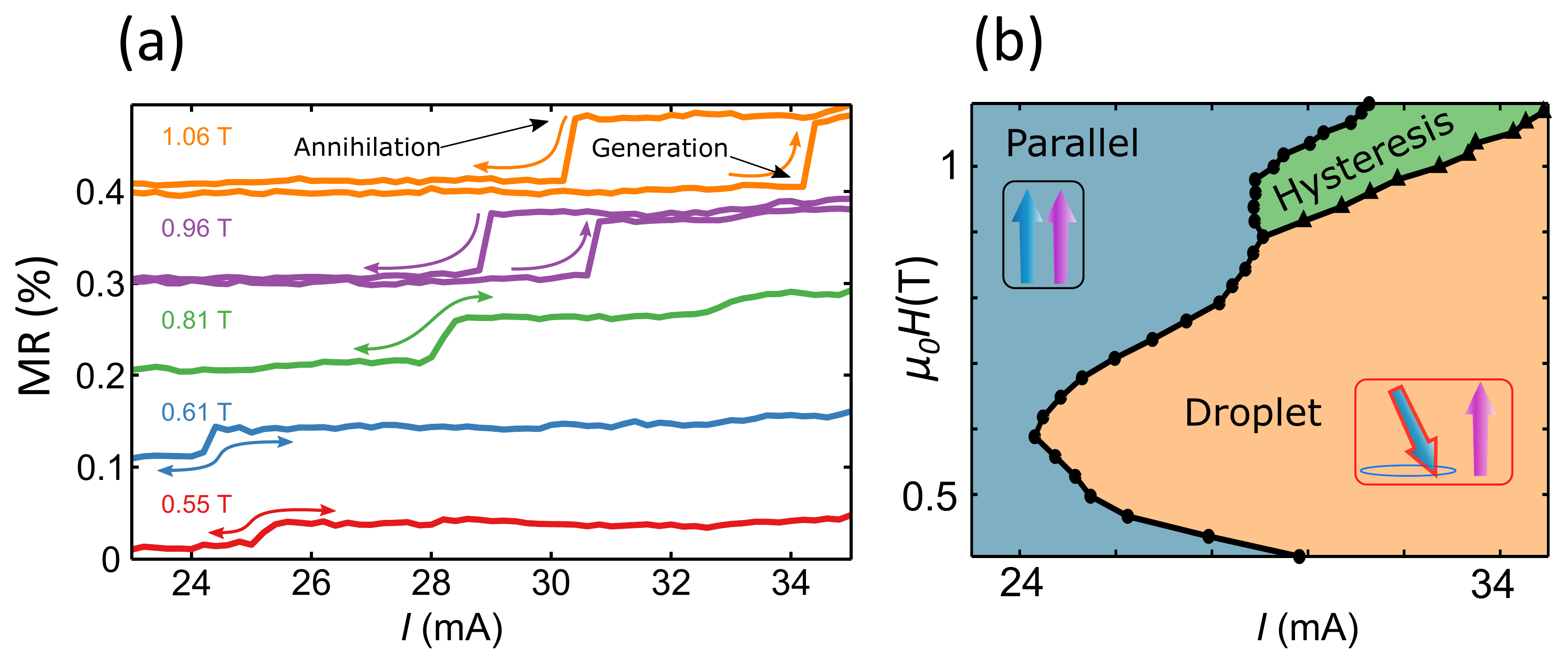}
		\caption{{\small (a) Magnetoresistance as a function of applied current for fields ranging from 0.55 to 1.06 T at room temperature. (b) Droplet boundary map. In the hysteretic area, triangles show droplet generation and dots annihilation.}}
		\label{Fig:diagram}	
	\end{center}
\end{figure*}

A droplet boundary map in field and current can be made from magnetoresistance measurements. Figure~\ref{Fig:diagram}(a) shows the nanocontact resistance as a function of current at different fields where generation---resistance step up---and annihilation---resistance step down---characteristics are visible. Figure~\ref{Fig:diagram}(b) shows the droplet boundary map with both the threshold and annihilation currents for all measured fields. Hysteresis phenomena is observed only above 0.9 T. Lower temperature measurements showed a wider field and current hysteresis~\cite{Macia2014,Lendinez2017}.

The boundary map shows how the threshold current for droplet generation depends on applied field.
At fields sufficient to saturate the PL magnetization the threshold current is proportional to the applied field~\cite{Hoefer2010}. This is expected as the spin precession frequency is proportional to the applied field, as is the damping torque (which goes as $dm/dt$). Larger spin torques and thus currents are required to overcome the damping and nucleate a droplet as the field increases. However, at fields that do not completely align the PL magnetization with that of the FL, the degree of spin polarization of the electrical current in the direction of the FL plays an important role. This component of spin polarization depends on the magnetization component of the PL normal to the film plane, and the magnetization is proportional to the applied field. Thus, the threshold current decreases with increasing applied magnetic field at low fields, as reported in several experimental studies \cite{Mohseni2014,Macia2014,Chung2016,Ozyilmaz2003}, and shown in  Fig.~\ref{Fig:diagram} below 0.6 T. A simple model for these droplet map characteristics is presented and discussed in the supplementary material of Ref.~\cite{Macia2014}.

\section{Drift resonances and droplet modes}
\label{Sec:Drift}
Room-temperature measurements show that droplets form at an abrupt threshold in both current and field~\cite{Mohseni2014,Macia2014,Chung2016,Lendinez2015,MOHSENI_physicaB,Lendinez2017,Mohseni_2018}. However, the resistance jump when a droplet forms is usually much less than the change expected for a droplet with a fully reversed core; there has been evidence for full magnetization reversal beneath the nanocontact in resistance measurements at low temperature~\cite{Macia2014,Lendinez2017}. Room-temperature measurements showed much smaller hysteretic effects as well. Drift instabilities~\cite{Lendinez2015,Wills2016}---caused by asymmetries in the system such as variations of either effective field or magnetic anisotropy in the nanocontact region---create low-frequency dynamics (at hundreds of MHz) of the droplet that can explain both the lower magnetoresistance signal and the small hysteresis.

\begin{figure}[t]
	\begin{center}
		\includegraphics[width=\columnwidth]{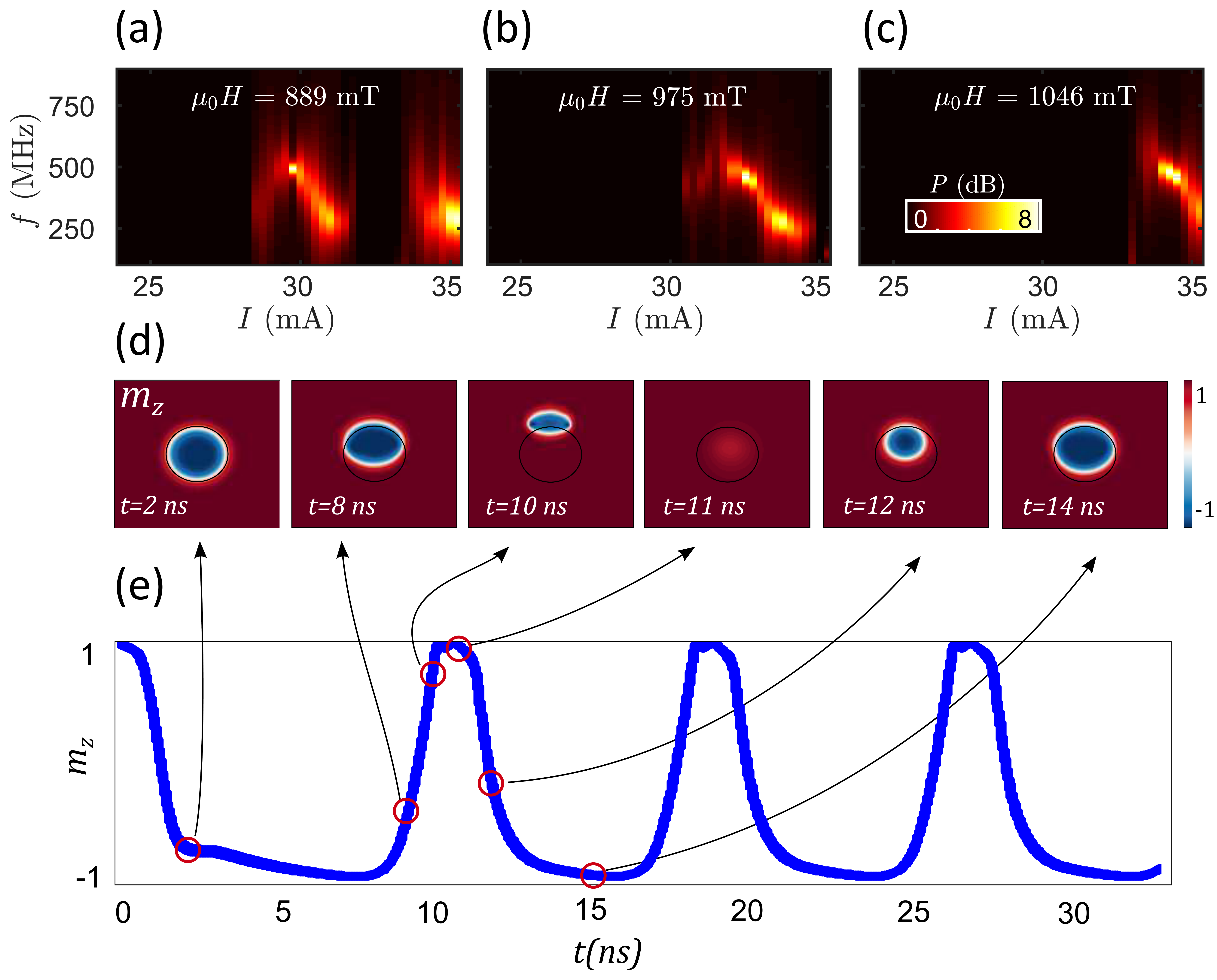}
		\caption{{\small Low-frequency spectra, color scale in dB above the baseline noise, as a function of applied current for fields of (a)  889 mT, (b) 975 mT, and (c) 1046 mT. In (d) and (e) micromagnetic simulation of the  evolution of a droplet soliton in an applied field of 1.1 T perpendicular to the film plane and an in-plane field ($x$ direction) of 0.15 T. Panels in (d) show magnetization maps of $m_z$ at particular times of the simulation. Images correspond to a 400 $\times$ 400 nm$^2$ field of view. The nanocontact region is outlined in black. In (e) the time evolution of the perpendicular component of the magnetization $m_z$ averaged over the nanocontact area is shown.}}
		\label{Fig:LFnoise}	
	\end{center}
\end{figure}

The low-frequency dynamics is associated with the collective motion or deformation of the droplet. While the spins precession in the droplet is at much higher frequency ($\sim 20$ GHz). The low-frequency dynamics can be detected electrically because the overall nanocontact resistance depends on the fraction of the droplet (reversed spins) that are exactly beneath the nanocontact area. This  Figures~\ref{Fig:LFnoise}(a-c) show measurements of nanocontact voltage noise at low frequencies during a current sweep. The low-frequency signal appears after the step increase in resistance (as in Fig.~ \ref{Fig:diagram}(a)) and is associated with the creation of the droplet excitation. There is a strong and broad oscillating signal at about 300 MHz with a weak dependence on applied field and current. This low-frequency noise has been found in many different experiments  \cite{Lendinez2015,Chung2016,Lendinez2017,Chung2018,Statuto2018} and has been associated with droplet drift instabilities.

Micromagnetic modeling shows that a small in-plane field (or a variation in the effective field within the nanocontact) causes a droplet to shift in the direction perpendicular to the applied in-plane field and annihilate as it moves outside the nanocontact. It annihilates because outside the nanocontact region there is damping that is not compensated by spin torque. Immediately after this, a new droplet is created beneath the nanocontact. Figure\ \ref{Fig:LFnoise}(d) and (e) show the evolution of a droplet in an
applied field of 1.1 T perpendicular to the film plane with an additional in-plane field of 0.15 T applied at $t =5$ ns. A droplet forms first with all spins precessing in phase. At $t >$ 5 ns the in-plane field is applied in the $x$ direction creating a drift instability, an imbalance in the precession phases that shifts the droplet in the $y$ direction (perpendicular to the applied field) until it annihilates. At $t = 10$ ns the first droplet has dissipated but a new one is being created. The time average change of $m_z$ beneath the nanocontact---that is the measurable quantity using any dc technique---is largely reduced, in this case to only 36 \% of the total.

The nanocontact temperature will be a few tens of degrees higher than the rest of the sample due to the high local current density~\cite{Petit-Watelot2012}. Magnetoresistance and electrical noise measurements have shown that droplets become more stable at lower temperature \cite{Lendinez2017} and have lower current density thresholds, in contrast to typical spin-transfer-torque-induced switching devices in which the switching threshold increases with decreasing temperature. The reason for this is not clear but may indicate that thermally excited (incoherent) magnons inhibit droplet formation.

The stability/instability of droplets remains a matter of study given their complexity as a nonlinear wave structure in systems with considerable thermal noise \cite{Wills2016,Chaves2020}. Experimental studies have suggested periodic deformations of its perimeter~\cite{Mohseni2014,puliafito2014self,Xiao_prb2017} and also multiple and, under certain conditions, combinations of droplet modes, each with a distinct high-frequency spin precession~\cite{Statuto2018}.

\section{Generation and annihilation timescales}
\label{Sec:timescales}
Basic questions about droplet characteristics are related to the timescales for their generation and annihilation.  That is, once a current (and associated spin-transfer torque) is present how long does it take to form a droplet soliton. A related question---and one important for consideration of these objects as mobile information carriers---is once formed how long does a droplet persist in the absence of a sustaining current, for example, outside a nanocontact region. 

\begin{figure}[t]
	\begin{center}
		\includegraphics[width=0.9\linewidth]{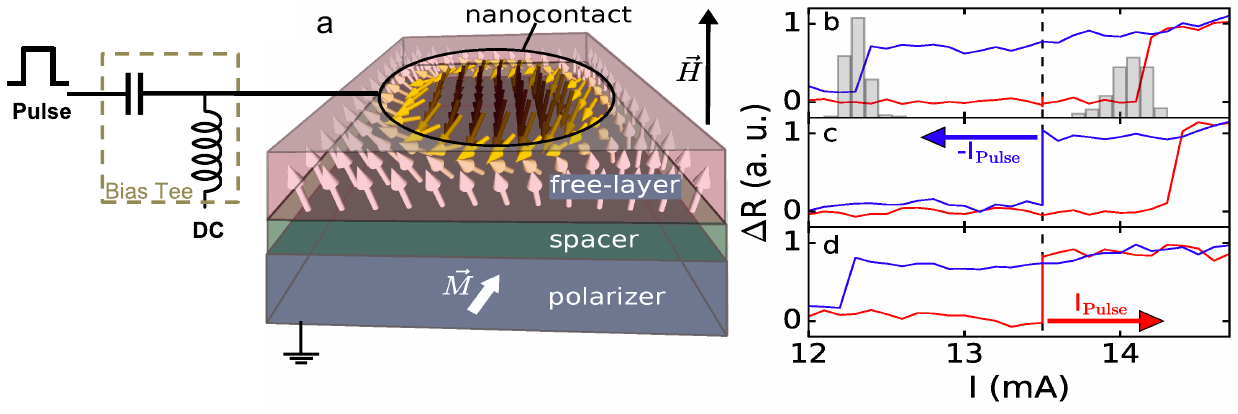}
		\caption{(a) A nanocontact is biased with a dc current in a hysteretic region, a region in which both droplet and non-droplet states are possible, as seen in the resistance versus current data shown in (b). A pulse is then applied to bring the nanocontact momentarily outside this region. (c) A negative current pulse annihilates the droplet, as seen by the step decrease in nanocontact resistance. (d) Starting in the non-droplet state a positive current pulse can generate the droplet, as seen by the step increase in nanocontact resistance. From ref.~\cite{Hang2018}}
		\label{Fig:GenAnn}	
	\end{center}
\end{figure}

\begin{figure}[t]
	\begin{center}
		\includegraphics[width=0.9\linewidth]{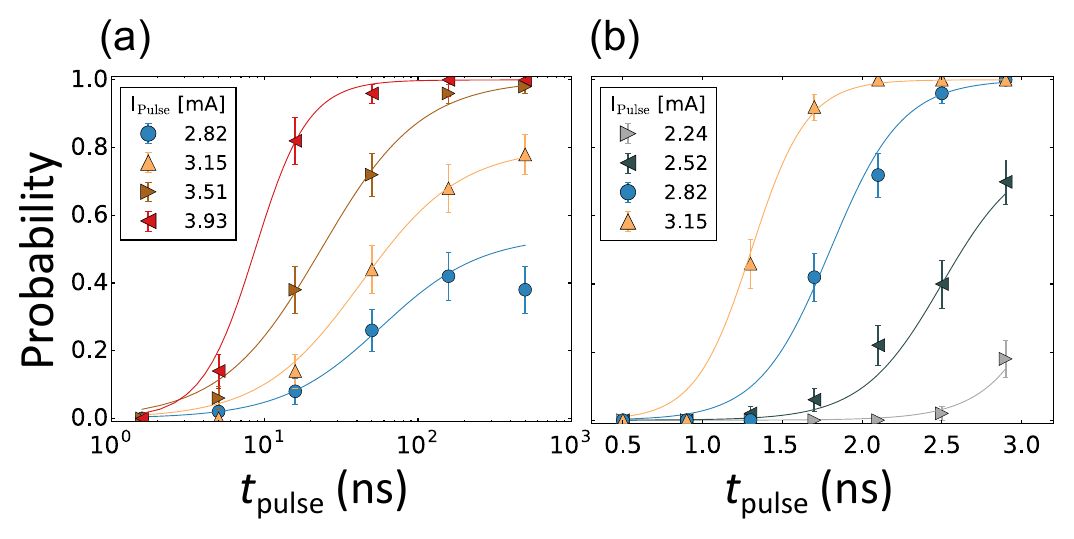}
		\caption{a) The generation probability versus applied pulse duration for different pulse amplitudes. Time on the x-axis is plotted on a logarithmic scale up to $1 \;\mu$s. b) The annihilation probability versus pulse duration again for different pulse amplitudes. The x-axis is now a linear scale with a maximum pulse duration of 3 ns. In both cases the dc current was fixed at 13.5 mA and a 0.7 T field was applied. From ref.~\cite{Hang2018}}.
		\label{Fig:Timescales}	
	\end{center}
\end{figure}

The droplet hysteretic response as a function of current described earlier (in Sec.~\ref{Sec:GenAnn}) enables experiments to determine these timescales. The basic idea of the experiment is shown in Fig.~\ref{Fig:GenAnn}. 
A nanocontact is biased at a current in a hysteretic region.
The experiments determine the probability of nucleating a droplet starting from a non-droplet state and, conversely, the probability annihilating a droplet starting from a droplet state. The results are shown in Fig.~\ref{Fig:Timescales}. Figure~\ref{Fig:Timescales}(a) shows that it can take 10 to 100s of nanoseconds to nucleate a droplet while droplets are annihilated in just a few ns for the same range of pulse amplitudes. This suggests that different processes are involved in droplet generation and annihilation.

Generation involves exciting a sufficient number of magnons to form a droplet. As each electron transmitted through the nanocontact excites at most one magnon the nucleation time will depend on the total charge flow through the nanocontact. A simple order of magnitude estimate is $(I-I_{c0})P t_\mathrm{pulse}=N_m$, where $N_m$ is the number of magnons required to form a droplet and $P$ is the spin polarization. For a 100 nm diameter nanocontact and a 3 nm thick layer, $N_m \simeq MV/\mu_B = 10^6$, where $M$ is the layer magnetization $7.5\times10^5$ A/m and $V$ is the droplet volume. With $(I-I_{c0})=3$ mA and $P=0.1$ one finds $t_\mathrm{pulse}=1$ ns.  Experiments give an order of magnitude longer time scales.

Micromagnetic model of the droplet formation shows longer generation times scales as well; in simulation this is associated with a delayed start to the droplet formation, what has been termed an incubation delay~\cite{Devolder2008}. The delay is associated with very small transfer of angular momentum in the initial (equilibrium) magnetization state, when the spin polarization is nearly collinear with the magnetization~\cite{Hang2018}. In the simple estimate this can be characterized by $P\ll1$ at the onset of the current pulse.

It is interesting to compare the droplet annihilation time to average magnon relaxation time, $\tau_m=1/(\alpha \omega)$, where $\omega$ is the angular spin precession frequency. With $\omega=2\pi \;2\times10^{10}$ rad/s and $\alpha = 0.03$, $\tau_m=0.2$ ns. This is about an order of magnitude smaller than the observed droplet decay time. In fact, micromagnetic simulations show much longer timescales for droplet `evaporation.' A faster decay mode are droplet drift resonances in which the droplet drifts outside the contact and is no longer sustained by a spin-transfer torque, as discussed in Sec.~\ref{Sec:Drift}~\cite{Hang2018}. The pulse experiments do not distinguish between a droplet evaporating in the contact and one drifting outside the contact region.

\section{Perspectives}

Interest in droplets is associated with their characteristics as nanometer scale nonlinear oscillators with large frequency tunability. We note that droplet dynamics are important to several applications. Foremost, to the development of improved spin oscillators, $>20$ GHz current controlled oscillators with higher quality factor, coherence and output signals. Further, their hysteresis and low frequency dynamics may be useful for applications as neuromorphic devices. Remarkably, droplets have been found to be relevant to today's state-of-the-art perpendicular magnetic tunnel junctions, devices being very actively developed by the semiconductor industry~\cite{KentWorledge2015}. Micromagnetic studies have shown that spin-torque reversal can start with the formation of a droplet in the center of the tunnel junctions free layer which then experiences a drift instability leading to a domain wall that traverses the element~\cite{Mohammadi2020,Volvach2020}. These connections were both unexpected and highlight the often unanticipated connections between basic studies and applications.

\subsection{Droplets Oscillators}

Spin waves are localized to the contact region and spin-wave radiation from the contact is minimized. As a result this channel of energy loss is reduced if not eliminated entirely. Thus one can expect that droplets would form high quality factor oscillators. To date the reported values are around $Q=2000$ \cite{Lendinez2015,Statuto2018}, which are similar to the obtained with spin torque oscillators based on easy plane ferromagnetic layers, including magnetic tunnel junctions~\cite{SILVA20081260,joovon2012,Chen2016}.

Spin torque vortex oscillators are also based on localized magnetic excitations~\cite{Pribiag2007} with minimal spin-wave radiation but with oscillations frequencies usually less than a few GHz (compared to droplets with 10s of GHz oscillation frequencies). It is not clear whether droplet could form even higher quality oscillators. Perhaps, if droplets could be rendered stable at room temperaure in a nanocontact or confined to the nanocontact region, e.g., by film geometry, their quality factor could be increased \cite{Wills2016}. The measurement of linewidth is sometimes insufficient to determine the quality of an oscillator given their strong nonlinearity that results in movement of the central frequency~\cite{Keller_2010_noise}.

Pure spin currents can also be used to generate spin wave excitations in magnetic nanostructured by the spin Hall effect \cite{Demidov2012,demidov2017}. Spin Hall nanooscillators do not require flow of electrical charges through the magnetic layer, which causes Joule heating, electromigration effects and Zhang-Li torques. It is also possible to use insulating magnetic materials in spin Hall nanooscillators~\cite{Xiao2012}. Droplet soliton modes have been created by pure spin currents in nanoconstriction-based spin Hall devices~\cite{Divinskiy2017,Chen2020}.

Another important characteristic of an oscillator is its output signal. The signal is directly related to the contact's magnetoresistance. Metallic structures have much smaller magnetoresistance than magnetic tunnel junctions~\cite{Maehara_2014} and thus droplet oscillators based on MTJs could have much larger output powers. Tunnel junction based nanocontacts have been fabricated but thus far have not shown evidence for droplets~\cite{Maehara_2014,Jiang2019}.

Synchronizing multiple oscillators provides a means to improve their quality factor and output signal. Coupled spin-Hall oscillators have proven high quality factors at high frequencies up to $Q=170000$ in the case of an array of 64 synchronized oscillators~\cite{Zahedinejad2020}. Oscillators can be coupled electrically (e.g. placed in series or parallel) to enhance their output signal~\cite{Grollier2006} or magnetically, examples being magnetic dipole interactions or coupling by spin waves~\cite{Kaka2005,Mancoff2006}. The latter coupling mechanism is limited for droplets because spin waves are localized. In an array of oscillators this could be a feature rather than a drawback. A network of STNO could couple below the threshold for droplet formation (at currents in which the nanocontact emits spin waves) and formation of a droplet could be used to controllably decouple oscillators. Droplets may also phase lock to an oscillating microwave magnetic field~\cite{Liu_phaselock2017}. Another interesting feature is that droplets have a particle-like nature, e.g. drift resonances and breathing modes, that could be used to increase the complexity of their interaction, including, for example,  synchronization to a low frequency background.

\subsection{Droplets and Bullets}
\begin{figure}[t]
	\begin{center}
		\includegraphics[width=0.9\linewidth]{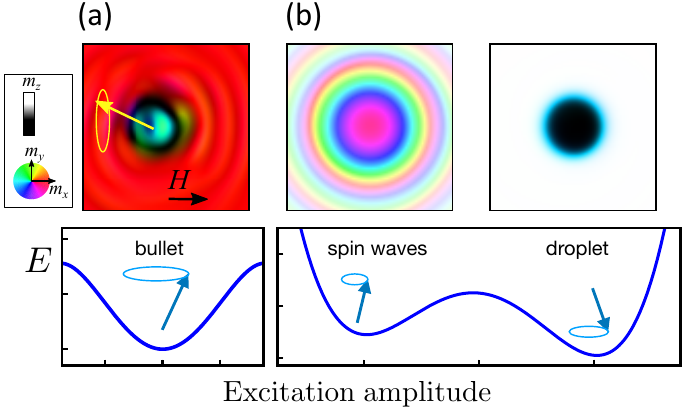}
		\caption{(a) A spin wave bullet corresponds to a large amplitude localized excitation in an easy plane magnetic layer. The excitation energy increases monotonically with spin-precession amplitude. (b) In an easy axis magnetic layer spin waves radiate away from a nanocontact at low current with frequencies close to the film's FMR frequency. Larger currents lead to droplet formation with localized spin waves. An energy barrier separates spin wave and droplet states leading to a hysteretic response to current and applied field.}
		\label{Fig:BulletsDroplets}	
	\end{center}
\end{figure}

Droplets have some characteristics similar to what has been termed a spin-wave bullet~\cite{Slavin2005} in that both involve large amplitude magnetic excitations and can be localized modes~(Fig.~\ref{Fig:BulletsDroplets}).
 There are important differences as well. Spin-wave bullets occur in magnetic layers with an easy-plane magnetic anisotropy, whereas droplets form in materials with an easy axis type anisotropy. Further, bullets result from a continuous transition to a localized spin-wave mode with increasing mode amplitude and decreasing spin-precession frequency (a red shift) with increasing current. For droplets there is an energy barrier to their formation and, once formed, to their annihilation. No such barrier appears to exist for bullet modes. Their similarity is that both are induced by a spin current typically in a nanostructure or nanocontact. Spin-Hall effect oscillators have thus far been of the ``bullet'' type, as the in-plane polarized spins from the spin-Hall effect are much more effective in exciting spin waves in in-plane magnetized materials. The recently reported spin currents associated with the planar Hall effect~\cite{Safranski2020} may provide a means to create droplets, as the spin currents can have a polarization perpendicular to the film plane.

\subsection{Droplet Merging and Propagation}

Manipulation of droplets, including propagation and interaction among them, is of importance for their applications. Droplets, as particle-like objects, can propagate and interact with each other. Droplets have an additional degree of freedom---their phase---associated with their precessional nature, which makes droplet interactions very interesting. Interaction behaviors have been classified through micromagnetic simulations~\cite{Bookman2014} and include repulsive and attractive interactions causing either droplet merging or droplet reflection.

Experimentally, a driving force needs to be applied on the droplet to induce a first drift---it has been shown that an effective field gradient can do this job~\cite{Lendinez2015}. Next, one would need a lower damping material to maximize the distance a droplet can travel before it annihilates~\cite{Bookman2014}. Studies have been conducted of merging two individual current driven droplets into a single droplet \cite{Xiao_prb_2016,Wang_AIP_merging2018}. They have considered the action of the external applied field---including field pulses---together with the applied current values together with the droplet sizes in this process.

\section{Summary}
This review has highlighted some of the basic characteristics of droplet solitons generated by spin transfer torques and x-ray microscopy imaging and electrical studies that have provided great insight into their dynamics. It is exciting that ideas about magnetic solitons in perpendicularly magnetized thin films that were explored theoretically in the 1970s in an idealized setting (a magnetic material without damping) can now be studied experimentally in great detail, so far, in transition metal multilayers. In the future droplets may be formed and explored in magnetic insulators using spin-orbit torques and similar physics may be found in antiferromagnets exited by spin torques. Experiments on droplets have also led to a deeper theoretical understanding of their dynamics and stability.

\section*{Acknowledgements}
We thank our many collaborators over the years that have contributed to the experiments and ideas that have been discussed in this review: J. Albert, S. Bonetti, R. Kukreja, J. Hang, C. Hahn, J. M. Hern\`andez, S. Lend\'inez, H. Ohldag and N. Statuto. We thank Dan Stein for discussions and comments on this manuscript. We also thank Julie Grollier, Danijela Markovic and Mark Stiles for comments on this manuscript.
FM acknowledges support from Spanish government through Grants No. RYC-2014-16515, No. MAT2015-69144-P, No. SEV-2015-0496 and No.
MAT2017-85232-R. ADK received support to conduct the x-ray microscopy and droplet lifetime experiments from the National Science Foundation under Grant No. DMR-1610416. Research at NYU related to neuromorphic computing was supported by Quantum-Materials for Energy Efficient Neuromorphic-Computing, an Energy Frontier Research Center funded by the U.S. Department of Energy (DOE), Office of Science, Basic Energy Sciences (BES), under Award DE-SC001927.

The data that support the findings of this study are available from the corresponding author upon reasonable request.

\section*{Data Availability Statement}
The data that support the findings of this study are available from the corresponding author upon reasonable request.

\bibliography{DropletReferences} 
\end{document}